\documentstyle[12pt]{article}
\def\setb@se#1{\baselineskip=#1 \normalbaselineskip=#1}
\setb@se{14pt}
\newcommand{\be}{\begin{equation}}
\newcommand{\ee}{\end{equation}}
\newcommand{\ii}{{\it i}}

\begin{document}

\begin{titlepage}
\begin{flushright}
Z\"urich University Preprint

ZU-TH 27/94

August 1994

hep-th/9409041
\end{flushright}

\vspace{20 mm}
\begin{center}
{\Huge \bf Odd--parity negative modes \\

of Einstein--Yang--Mills black holes \\

and sphalerons \\

}

\vspace{10 mm}
\end{center}
\begin{center}
{\bf Mikhail S. Volkov}\footnote{On leave from Physical--Technical
Institute of the  Academy  of  Sciences of Russia, Kazan  420029,
Russia}
\vskip2mm
{\em Institut    f\"ur  Theoretische  Physik  der  Universit\"at
Z\"urich--Irchel, \\
Winterthurerstrasse 190, CH--8057 Z\"urich,
Switzerland,\\
e--mail: volkov@physik.unizh.ch}
\vskip1cm
{\bf Dmitri V. Gal'tsov}\footnote{On leave from
the Dept.\ of Theoretical Physics, Moscow State
University, 119899 Moscow, Russia; e--mail: galtsov@grg.phys.msu.su}
\vskip2mm
{\em Centro de Investigaci\'on y de Estudios Avanzados del I.P.N.,\\
Departamento de F\'\i sica,
Apdo. Postal 14-740, 07000,
M\'exico, D.F., M\'exico,\\  e--mail: dgaltsov@fis.cinvestav.mx}
\end{center}
\vspace{20mm}\begin{center}{\bf Abstract}\end{center}
\vspace{2mm}

An analytical proof  of  the  existence  of  negative  modes  in  the
odd--parity  perturbation  sector is given for all known non-Abelian
Einstein--Yang--Mills
black holes. The significance of the  normalizability  condition  in
the functional stability  analysis is emphasized. The  role  of the
odd--parity negative modes in the sphaleron  interpretation  of  the
Bartnik--McKinnon solutions is discussed.

\end{titlepage}
\newpage

\section {Introduction}

Soon after  the  discovery   of   regular  particle--like  \cite{BK}
and   black   hole \cite{BH} solutions in the  $SU(2)$
Einstein--Yang--Mills (EYM)  theory,
their instability  was  demonstrated
\cite{SZ}. Numerical analysis has revealed the existence of $n$ (the
number of nodes  of  the  YM  function)  negative   modes   in   the
{\em even--parity} spherical perturbation  sector;  moreover,   the
investigation of the non-linear dynamics of perturbations has clearly
shown the instability \cite{Z} (an interesting global  analysis
can also be found in \cite{Ma}). Some time later, a generalization
of the $SU(2)$ black hole solutions for the non-vanishing charge was
found in the EYM theory
for a larger gauge group \cite{charge}, however, the issue of
stability for these solutions has been open so far.

It is worth noting that,  for  the  regular  Bartnik--McKinnon  (BK)
solutions,  there  exists  at  least  one  negative  mode   in   the
{\em odd--parity} sector too. This mode has the  same nature  as
negative mode of the electroweak sphaleron solution \cite{guys},
\cite{kunz}, indicating  the
presence of a potential barrier separating topologically  distinct
YM vacua. This plays the crucial role  in  the  proposed   sphaleron
interpretation for the BK  solutions  \cite{spha}, and  the  related
physical issues \cite{sphal}. The existence of the odd--parity
negative mode has been shown analytically using
variational techniques
\cite{spha}. The idea consists in the construction
of an   energy--reducing  function sequence  on the
constraint surface in configuration space  of  the   theory.
The advantage of this technique  is  that  it  does   not    require
the  detailed   knowledge    of    the    background     equilibrium
solution. Recently  this   approach has been successfully applied in
the  analysis  of   stability   of   regular   solutions   of   the
EYM--Higgs   theory  \cite{grand},   and   regular   solutions  of
the  EYM  theory  for an arbitrary gauge group \cite{brod}. The
latter result shows, in particular, the instability of the non-trivial
$SU(3)$ EYM regular solutions found recently by K\"unzle \cite{kunzle}.

For EYM black holes, the variational approach can  be   applied   as
well. For the $SU(2)$ solutions,
a sequence of trial functions reducing the  total  ADM  energy
has been constructed in \cite{old}. More detailed analysis  reveals,
however, that the decrease of the energy is not  sufficient
to  demonstrate  instability.   Another  important  condition
is  necessary:  the   energy    reducing perturbations   must   be
normalizable. Unfortunately, for the trial  functions  used   in  our
earlier paper \cite{old},   the   normalizability    condition   was
not satisfied because of their inappropriate behaviour
at  the  event horizon (for the regular  solutions  such  a  problem
does  not  arise). Surprisingly enough, we find that  a
similar situation can be often met in other papers, where the argument
based on the existence of the energy reducing fluctuations  is  used
--- little  attention  is  usually  paid  to   the   normalizability
condition.

The purpose of this paper is two--fold: first, we want  to
elucidate  the nature  and  to emphasize the  importance  of  the
normalizability condition for the general field--theoretical
stability analysis.
Secondly, we investigate the structure of the {\em odd--parity}
spherical perturbation sector for the static EYM black hole and
sphaleron solutions.
We  shall  give a
proof of the existence  of negative  modes  in   this
sector for all known essentially non-Abelian
EYM black holes, and our proof is valid in the regular case as well.
This justifies  the  main statement  made in \cite{old}
concerning the $SU(2)$ EYM black holes.
In addition, we analyse stability of magnetic
$U(1)$ black holes within the context of  EYM theory, and discuss
the importance of the odd--parity negative modes for
understading the physical nature of the localized finite energy
EYM solutions.
\section{The functional criteria of instability}

Consider a theory of  a   classical   field,   or   several  fields,
denoted commonly by  $\phi$.  Assume  that the   theory   admits   a
localized   static   equilibrium   solution    $\phi_{s}({\bf   x})$
possessing finite  energy $E[\phi_{s}({\bf x})]$.
Suppose one has found  a
sequence  of   static  field  configurations,   $\phi_{\lambda}({\bf
x})$,  such  that $\phi_{\lambda=0}({\bf x})   =\phi_{s}({\bf  x})$,
and   the   energy has a maximum for $\lambda=0$:
\be
E[\phi_{\lambda\neq 0}({\bf x})]<
E[\phi_{s}({\bf x})].                              \label{1}
\ee
The mechanical picture for this situation is that of a particle
sitting at the top of potential  hill  whose  profile  is  given  by
$U(\lambda)=  E[\phi_{\lambda}({\bf  x})]$. The question arises
whether (\ref{1}) is sufficient in order to  reveal an instability of
the solution $\phi_{s}({\bf x})$. Strictly  speaking,   the   answer
is negative.
Namely, the equilibrium state of the ``particle'' will be  unstable
only  if  its effective mass is finite.
Otherwise, the particle can not depart
from the top due to infinite inertia.

Let us reformulate this as  follows.  If  one  allows  for  a   time
evolution along the sequence  $\phi_{\lambda}({\bf  x})$,  replacing
the parameter $\lambda$ by a function $\lambda(t)$, and if one finds
that
the  kinetic energy, i.e. the part of the total energy  proportional
to  $\dot{\lambda}^{2}$ , is finite,  then   (\ref{1})   implies
indeed an
instability. Otherwise,  $\phi_{\lambda}({\bf x})$   is   not
a physically acceptable sequence, and nothing can be inferred from
Eq.(\ref{1}).

To put this into  more rigorous form,  consider  small  fluctuations
$\delta \phi(t,{\bf x})$   around   the  static equilibrium solution.
Linearizing the field equations and specifying the  time  dependence
as  $\delta\phi(t,{\bf x})= \exp(-i\omega t)\Psi({\bf x})$  one  can
usually represent the perturbation equations as \cite{grand}
\be
H\Psi = \omega^{2}M\Psi.                                 \label{2}
\ee
Here, the two operators $H$ and $M$ depending on $\phi_{s}({\bf x})$
are assumed to be independent on $\omega$. $H$  is  usually
self-adjoint with respect  to  a  properly  defined  scalar  product
$\langle\Psi|\Phi\rangle$,  and  the kinetic  energy  matrix $M$ is
positive definite, $\langle\Psi|M|\Psi\rangle  > 0$.

Clearly, the equilibrium static solution  will  be  unstable  if   a
normalizable solution to (\ref{2})   with    $\omega^{2}<0$    can
be found. However, this may require numerical calculations.
The  simpler  (though  less  informative)  way  to   reveal
an instability is  to  make  use  of  the  minimum   principle   for
the following functional defined through  (\ref{2}):
\be
\omega^{2}(\Psi)=\frac{\langle\Psi|H|\Psi\rangle}
{\langle\Psi|M|\Psi\rangle},                             \label{3}
\ee
with $\Psi$  being a trial function. The lowest eigenvalue is  known
to correspond to the lower bound for this functional.  From  here it
follows   that   if a   $\Psi$   can    be    found  such  that
$\omega^{2}(\Psi)<0$,  then  the  operator  $H$  is   not   positive
definite, and negative eigenvalues therefore do exist.

Obviously, $\omega^{2}(\Psi)<0$ implies that
\be
\langle\Psi|H|\Psi\rangle~<0,                             \label{4}
\ee
which is equivalent to (\ref{1}).  This  condition is  frequently
used   in   order    to    demonstrate instability    in
field--theoretical systems (see for instance \cite{guys}, \cite{hol}).
However, for such systems the operator $M$ is generally  unbounded.
Often, the trial functions from the domain
of $H$, while ensuring (4), lead  to  {\em divergence}
of  the expectation value
$\langle\Psi|M|\Psi\rangle$. In  this  case (\ref{4})
 is   insufficient   to
establish instability. We therefore arrive at the  second  important
condition
\be
\langle\Psi|M|\Psi\rangle~<\infty.                        \label{5}
\ee
Physically this can be seen as the condition assuring the finiteness
of  the  ``kinetic   energy''   associated   with    time--dependent
perturbations.

We deliberately spend so much time on these subtleties since in  the
existing literature they seem  often  to  be  overlooked.  Sometimes
the trial functions   used   automatically fulfill  condition
(\ref{5}), but  not  always.  One  may think  that  condition
(\ref{5}) is  somehow  implied by (\ref{4}), in the  sense that,  if
a function subject to  (\ref{4}) is known,  then    one    can    in
principle    find       another   function     satisfying       both
(\ref{4})   and   (\ref{5}).  There are,  however,   examples   when
condition  (\ref{4}) holds   even   for   perfectly    stable
solutions   ---   but    not both     (\ref{4})     and   (\ref{5}).
Consider, for instance, the Bogomolny--Prasad--Sommerfield  monopole
solutions  \cite{BPS}. They   form    a    family   $\phi_{BPS}({\bf
x},v)$  whose  members are  distinguished  by  the asymptotic  value
of the  Higgs  field, $v$,   and   satisfy   the   same  system   of
equations;  the corresponding  mass  is proportional  to $v$.  Given
a member of the family  specified  by   some   $v_{0}$,    one   can
construct  the      trial          sequence     $\phi_{\lambda}({\bf
x})=\phi_{BPS}({\bf  x},v_{0}-\lambda^{2})$  which    reduces    the
energy.  However,  the  norm    (\ref{5})    is   divergent  due  to
the change of the asymptotic values of the fields.

Finally, we sketch an idea of the rigorous justification of the above
functional criteria for instability. Assume that the matrix $M$ in
(\ref{2}) is non-degenerate and symmetric.
Perform  a  linear
transformation      $\Psi=O\tilde{      \Psi}$       such       that
$O^{T}MO=1$    and    the     new     Hamiltonian
$\tilde{H}=O^{T}HO$. Omitting  tilde, one can represent  (\ref{2})
as an eigenvalue problem, $H\Psi=\omega^{2}\Psi$, for the  unbounded
(but usually bounded from below)
operator $H$ acting on  the  Hilbert  space  ${\cal  H}$  of  $\Psi$
with  a  scalar  product   $\langle\Psi|\Phi\rangle   =\int
\Psi^{\dagger}\Phi~d\mu({\bf x})$, where $d\mu$  is an appropriately
chosen measure. Usually, one can easily specify a dense  set  ${\cal
D}(H) \subset {\cal H}$ as the domain of $H$ in
such a way that $H$ is symmetric on ${\cal D}(H)$. Next, assume that
{\it the existence} of a self-adjoint extension $H_{1}$ is known,
${\cal D}(H_{1})\supset{\cal D}(H)$. Then, the spectral decomposition
for $H_{1}$ implies the following inequality for the ground state
eigenvalue:
\be
\omega^{2}_{0} \leq\omega^2(\Psi),\ \ \
{\rm where}\ \ \omega^{2}(\Psi)=
\frac{\langle\Psi|H_{1}|\Psi\rangle}
{\langle\Psi|\Psi\rangle}, \ \ \ \Psi\in{\cal D}(H_{1}).
                              \label{5:1}
\ee
Thus, if $\omega^{2}(\Psi) < 0$ for  some  $\Psi\in{\cal  D}(H_{1})$,
then $\omega_{0}^{2}<0$. Notice that if one chooses  $\Psi\in  {\cal
D}(H)\subset{\cal D}(H_{1})$ then one can replace  $H_{1}$  by  $H$
in the definition of $\omega^{2}(\Psi)$.
The condition $\omega^{2}(\Psi)<0$  is  then  equivalent  to  the
following two conditions. The first one is given by (\ref{4}), where
the trial function $\Psi$ has to be an element of the already
specified set ${\cal D}(H)\subset{\cal H}$.  The   second,  the
normalizability condition, becomes now fairly trivial
\be
\langle\Psi|\Psi\rangle~<\infty;                         \label{5:2}
\ee
this simply states that $\Psi$ must belong to the Hilbert space
${\cal H}$.
Note that these arguments do not assume $H$ to be essentially
self-adjoint, which is often not easy to show. It suffices to
ensure that a self-adjoint extension for $H$ {\em exists},
for which one has simple powerful criteria \cite{RS}.

In what follows, we apply this procedure  within  the
context of EYM  theory.

\section{Existence of odd--parity negative modes for EYM black holes}

All known essentially non-Abelian EYM black
hole solutions \cite{BH}, \cite{charge} can be
obtained within the context of the $SU(2)\times U(1)$
EYM theory. The corresponding action is
\be
S=-{1\over 16\pi G}\int R\sqrt{-g}d^{4}x - \int\left(
{1\over 2e^{2}} trF_{\mu\nu}F^{\mu\nu}+
{1\over 4} {\cal F}_{\mu\nu}
{\cal F}^{\mu\nu}\right) \sqrt{-g}d^{4}x,                 \label{6}
\ee
where       $F_{\mu\nu}=\partial_{\mu}A_{\nu}-\partial_{\nu}A_{\mu}-
\ii [A_{\mu},A_{\nu}]$ is the matrix valued gauge   field    tensor,
$e$       is       the       gauge         coupling        constant,
$A_{\mu}=A_{\mu}^{a}\tau^{a}/2$, and  $\tau^{a}\   (a=1,2,3)$    are
the  Pauli  matrices; ${\cal F}_{\mu\nu} $ is the $U(1)$ field
strength.

In the spherically  symmetric   case,   the   spacetime   metric  is
chosen to be
\be
ds^{2} = l^{2}_{e}\left( (1-2m/r)\sigma^{2}dt^{2} -
\frac{dr^{2}}{1-2m/r} - r^{2}(d\vartheta^{2} + \sin^{2}\vartheta
d\varphi^{2})\right).                                     \label{7}
\ee
Here $l_{e}=\sqrt{4\pi}l_{pl}/e$ is the only dimensional quantity in
the problem ($l_{pl}$ being the Planck length); the functions $m$
and $\sigma$ depend on $t$ and $r$.
The $U(1)$ part of the gauge field is chosen to be of the dyon
type, $e{\cal F}=(q_{e}/r^{2})dt\wedge dr+
q_{m} \sin\vartheta~d\vartheta\wedge d\varphi$,
satisfying the Maxwell equations for any
constant $q_{e}$ and $q_{m}$.

The spherically symmetric $SU(2)$ YM field can be parameterized by
\be
eA = W_{0}\hat{L}_{1}\ dt + W_{1}\hat{L}_{1}\ dr
+\{p_{2}\ \hat{L}_{2} - (1-p_{1})\
\hat{L}_{3}\}\ d\vartheta
+\{(1-p_{1})\ \hat{L}_{2} + p_{2}\
\hat{L}_{3}\}\sin\vartheta\ d\varphi,
                                                        \label{8}
\ee
where  $W_{0}$, $W_{1}$, $p_{1}$, $p_{2}$ are functions of  $t$  and
$r$,           $\hat{L}_{1}           =            n^{a}\tau^{a}/2$,
$\hat{L}_{2}=\partial_{\vartheta}\hat{L}_{1}$,
$\sin\vartheta\hat{L}_{3}=\partial_{\varphi}\hat{L}_{1}$,        and
$n^{a}=(\sin\vartheta          \cos\varphi,\sin\vartheta\sin\varphi,
\cos\vartheta)$.  The  gauge  transformation
\be
A \rightarrow\  UAU^{-1}+ \ii UdU^{-1}, \ \ {\rm with}\ \
U=\exp(\ii\Omega(t,r)\hat{L}_{1}),                  \label{10}
\ee
preserves the form of the field (\ref{8}), altering the functions
$W_{0}$, $W_{1}$, $p_{1}$, $p_{2}$ as
\be
W_{0}\rightarrow W_{0}+\dot{\Omega},\ \ \
W_{1}\rightarrow W_{1}+\Omega',\ \ \
p_{\pm}=p_{1}\pm\ii   p_{2}
 \rightarrow \exp(\pm\ii\Omega)p_{\pm};        \label{11}
\ee
here dot and prime denote differentiation with respect  to  $t$  and
$r$, respectively.

It is convenient to  introduce  the  complex  variable  $f=p_{1}+\ii
p_{2}$ and its covariant derivative $D_{\mu}f =  (\partial_{\mu}-\ii
W_{\mu})f$,   as   well    as    the   $(1+1)$-dimensional     field
strength                     $W_{\mu\nu}=
\partial_{\mu}W_{\nu}-\partial_{\nu}W_{\mu}$,  $(\mu,   \nu  =0,1)$.
The full system of the EYM  equations then reads
\be
\partial_{\mu} (r^{2}\sigma W^{\mu\nu})-
2\sigma~ Im~ (fD^{\nu}f)^{\ast}=0,                         \label{12}
\ee
\be
D_{\mu}\sigma D^{\mu}f-\frac{\sigma}{r^{2}}(|f|^{2}-1)f=0, \label{13}
\ee
\be
m'=-\frac{r^{2}}{4}W_{\mu\nu}W^{\mu\nu} +
\frac{1}{N\sigma^{2}}|D_{0}f|^{2}+N|D_{1}f|^{2}+
\frac{1}{2r^{2}}(|f|^{2}-1)^{2}+\frac{q^{2}}{2r^{2}},      \label{14}
\ee
\be
\dot{m}=2N~Re~D_{0}f(D_{1}f)^{\ast},                       \label{15}
\ee
\be
(\ln\sigma)' = \frac{2}{r} \left(\frac{1}{N^{2}\sigma^{2}}
|D_{0}f|^{2}+|D_{1}f|^{2}\right),                         \label{16}
\ee
where $N=1-2m/r$, $q^{2}=q_{e}^{2}+q_{m}^{2}$,
and asterisk denotes complex conjugation.

Both for $q=0$ (the $SU(2)$ case) \cite{BH}, and $q\neq 0$
\cite{charge}, these equations are known to possess static non-Abelian
black  hole  solutions
\cite{BH} labeled by a pair $(n,r_{H})$, where $n$  is  an  integer,
and  $r_{H}\geq |q|$ is the black hole radius. For these  solutions,
$W_{0}=W_{1}= p_{2}=0$; the function $f=Re~f=p_{1}$ has $n$ nodes in
the domain $r_{H}<r<\infty$ and tends  asymptotically  to  $\pm   1$
always remaining in the stripe $-1<f(r)<1$.  The  metric   functions
$N$  and  $\sigma$  increase  monotonically  from  $N(r_{H})=0$   to
$N(\infty)=1$ and from  $0<\sigma(r_{H})<1$  to  $\sigma(\infty)=1$,
respectively. Asymptotically, the metric is of the
Reissner-Nordstr\"om (RN) form with charge $q$.

It is worth noting that $SU(2)\times U(1)$ group may also arise
as a subgroup of a larger gauge group \cite{charge}.
The basic EYM
equations remain of course the same in this case, and the only
difference may appear in the quantization condition for the Abelian
magnetic charge $q_{m}$.

Consider small perturbations of a given black hole solution
\be
m\rightarrow m+\delta m,\ \
\sigma\rightarrow\sigma +\delta\sigma,\ \
f\rightarrow f+\delta f,\ \
W_{0}=W_{1}=0\rightarrow \delta W_{0},\ \delta W_{1}.    \label{17}
\ee
The linearization  of   the   field   equation   reveals  that   the
even--parity   perturbations,   $(\delta   m,\delta\sigma,    \delta
p_{1})$,  and  the  odd--parity   perturbations,   $(\delta   W_{0},
\delta  W_{1}, \delta p_{2})$,  decouple, and
therefore are independent --- this is   because   the
background solutions are invariant under  parity.  Notice  that  the
infinitesimal  gauge    transformation    (\ref{10})    does     not
alter the even-parity perturbations, while the odd-parity ones change
as
\be
\delta W_{0}\rightarrow\delta W_{0}+\dot{\Omega}, \ \ \
\delta W_{1}\rightarrow\delta W_{1}+\Omega', \ \ \
\delta p_{2}\rightarrow\delta p_{2}+\Omega f.             \label{18}
\ee

For the {\em even-parity} modes, the perturbation equations  for
$(\delta m,\delta\sigma,  \delta   p_{1})$  can  be  reduced  to   a
single Schr\"odinger-type equation. In the $q=0$ case, this
equation was analyzed numerically by Straumann  and  Zhou  \cite{SZ}
who found exactly $n$ bound states for any background  $(n,  r_{H})$
black hole solution.
Here we will concentrate  on  the  {\em odd--parity}
sector, and our results do not depend on $q$.

For the odd-parity  fluctuations,  the  metric  remains  unperturbed
\cite{grand}, thus the perturbation equations can easily be
obtained via expanding the Yang-Mills equations (\ref{12}), (\ref{13})
with respect  to  $\delta  W_{0}$,  $\delta  W_{1}$,   and   $\delta
p_{2}$ alone. We use  the  gauge  freedom  (\ref{18})   to    ensure
the condition $\delta W_{0}=0$, and  specify  the  time   dependence
as  $\delta  W_{1}=\sqrt{2/(r^{2}N)}\alpha(r)\exp(-\ii\omega    t)$,
$\delta  p_{2}=\xi(r) \exp(-\ii \omega t)$. The resulting  equations
can   be  represented as the eigenvalue problem
\be
H\Psi=\omega^{2}\Psi, \label{18:1}
\ee
where $\Psi= \left(\begin{array}{c} \alpha \\ \xi
\end{array}\right)$,
and the Hamiltonian is given by
\be
H=\sigma N\left(\begin{array}{ccc}
2\sigma f^{2}/r^2 & & \sigma \sqrt{2N/r^{2}} (f'-if\hat{p})\\ \\
(f'+i\hat{p}f)\sigma\sqrt{2N/r^{2}} & &
\hat{p}\sigma N\hat{p} +
\sigma(f^{2}-1)/r^{2}\end{array}\right).                 \label{19}
\ee
The quantities $f, \sigma, N$ refer to the background static black
hole solution, and $\hat{p}=-id/dr$. There exists also an additional
equation due to the  Gauss  constraint  (Eq.(\ref{12})  with
$\nu=0$)
\be
\omega\left(\sqrt{\frac{2r^{2}}{\sigma^{2}N}}\alpha\right)'=
\omega\frac{2}{\sigma  N}f\xi.                    \label{20:1}
\ee
One  can see however that, as long as $\omega\neq 0$, this equation
is a  differential  consequence    of
(\ref{18:1}), (\ref{19}).  We   observe   therefore   that
any   solutions   to  the  eigenvalue  problem   (\ref{18:1})
(\ref{19}) with
$\omega\neq  0$ automatically satisfy the Gauss constraint, that is,
they have correct initial values.

We introduce the tortoise coordinate $r_{\ast}\in {\rm R}$, such that
$dr_{\ast}= dr/(\sigma N)$, and define the inner product as
\be
\langle\Psi_{1}|\Psi_{2}\rangle=\int_{-\infty}^{\infty}
\Psi_{1}^{\dagger}{\Psi_{2}}~dr_{\ast}=\int_{-\infty}^{\infty}
(\alpha_{1}^{\ast}\alpha_{2}+
\xi_{1}^{\ast}\xi_{2})dr_{\ast}.                   \label{19:1}
\ee
The Hilbert space can then be chosen to be
${\cal H}={\rm L}^{2}({\rm R},dr_{\ast})\oplus
{\rm L}^{2}({\rm R},dr_{\ast})$.
Consider the dense set in ${\cal H}$ consisting of twice
continuously differentiable functions with compact support. $H$
maps this set into ${\cal H}$ and it is symmetric on this set with
respect to the inner product (\ref{19:1}). We can therefore specify
this set as the domain of $H$, ${\cal D}(H)=
{\rm C}^{2}_{0}({\rm R})\times {\rm C}^{2}_{0}({\rm R})\subset
{\cal H}$ \cite{RS}, (in addition, $H$ can be shown to be bounded
from below on ${\cal D}(H)$).

Notice that $H$, being real, commutes with complex conjugation.
A theorem by Von  Neumann \cite{RS} then ensures the
existence of self-adjoint extensions for $H$. Thus,  according  to
the general procedure outlined above, to demonstrate instability  it
is sufficient to find a function  $\Psi\in{\cal D}(H)$
satisfying the inequality (\ref{4}).

Consider the set of real functions
$\{ h_{k}(r_{\ast})\} \subset{\rm C}^{2}_{0}({\rm R})$, where
$h_{k}(r_{\ast})=h(r_{\ast}/k)$, $k\geq 1$, and
$h(r_{\ast})$ is defined as follows:
$h(r_{\ast})=h(-r_{\ast})$, $h=1$ when $r_{\ast}\in [0,a]$,
$-D\leq h'_{\ast}<0$ for $r_{\ast}\in [a,a+1]$,
and $h=0$ for $r_{\ast}>a+1$. Here $a,D$ are positive constants,
prime and asterisk denote differentiation with respect to $r_{\ast}$,
$h'_{\ast}\equiv dh/dr_{\ast}=\sigma Nh'$.

Using $\{ h_{k}\} $, construct the set
$\{ \Psi_{k}\} \subset {\cal D}(H)$ specified by
\be
\alpha_{k}=\sqrt{2Nr^{2}}\sigma f'_{\ast}
\frac{(f^{2}-1)}{r^{2}}h_{k}\equiv\alpha_{0}h_{k},\ \ \ \
\xi_{k}=\left(f'_{\ast}\right)^{2}h_{k}\equiv\xi_{0}h_{k}, \label{21}
\ee
(notice that the background solutions $(f,f',N,\sigma)$ are at
least twice differentiable \cite{bfm}).
Integration by parts allows us to represent the expectation value
$\langle\Psi_{k}|H|\Psi_{k}\rangle$ in the form
\be
\langle H\rangle=
\int_{-\infty}^{\infty}
\left\{ \left( (\xi_{k})'_{\ast}-f\sigma N
\sqrt{\frac{2}{Nr^{2}}}\alpha_{k}\right)^{2}+
2\sigma N f'_{\ast}
\sqrt{\frac{2}{Nr^{2}}}\alpha_{k}\xi_{k}+
\frac{\sigma^{2}N}{r^{2}}(f^{2}-1)\xi_{k}^{2}\right\} dr_{\ast}.
                                                          \label{22}
\ee
Substituting (\ref{21}) into (\ref{22}) and taking into
account the background Yang-Mills equation
\be
f''_{\ast\ast}=\frac{\sigma^{2}N}{r^{2}}f(f^{2}-1)       \label{23}
\ee
we obtain
$$
\langle\Psi_{k}|H|\Psi_{k}\rangle
 =5\int_{-\infty}^{\infty}dr_{\ast}~\xi_{0}^{2}N\sigma^{2}
\frac{(f^{2}-1)}{r^{2}} +
$$
\be
\int_{-\infty}^{\infty}dr_{\ast}
\xi_{0}^{2}\left\{ (h_{k})'^{2}_{\ast}+
5\frac{N\sigma^{2}}{r^{2}}(1-f^{2})(1-h_{k}^{2})\right\}
\equiv I+I(k).                                         \label{24}
\ee
In this equation, the $k$--indepentent first term $I$ is finite and
manifestly negative (recall that  $|f|<1$, in addition,
$f'_{\ast}\sim N\rightarrow 0$
as $r_{\ast}\rightarrow -\infty$, $f'_{\ast}\sim 1/r_{\ast}^{2}$ as
$r_{\ast}\rightarrow \infty$, so the integral exists).
Taking into account the properties of the smoothening functions
$h_{k}$ specified above, one can see that the integral $I(k)$
converges uniformly and tends to zero as $k$ increases.
We therefore conclude that, for large $k$, the contribution
of $I(k)$ into (\ref{24}) is negligible, thus the whole expression
turns out to be negative.
This shows that
$\Psi_{k}$ with sufficiently large $k$ are such functions from
${\cal D}(H)$ that $\langle\Psi_{k}|H|\Psi_{k}\rangle<0$,
which completes the proof of the existence of the odd-parity negative
modes.
Physically, these $\Psi$ correspond to such admissible
time-symmetric initial data which give rise to a growing
instability \cite{wald}.

Our analysis is valid for any $r_{H}>|q|$.
For the uncharged $SU(2)$ solutions, the limit $r_{H}\rightarrow
0$ is known to relate to the regular case. Then $N(r)>0$ for any
$r\geq 0$, so $r_{\ast}$ runs over semi-axis, and the  Hilbert  space
is therefore ${\cal H}=
{\rm L}^{2}({\rm R}_{+},dr_{\ast})\oplus
{\rm L}^{2}({\rm R}_{+},dr_{\ast})$.
 One can see
that the analysis presented above is equally valid in this case.

Finally, for the sake of completeness, we analyse stability of the
Abelian magnetic $U(1)$ black holes. For these solutions,
the dynamics of perturbations is still governed by Eqs.(\ref{18:1}),
(\ref{19}), where the parameters of background solutions are
$W_{0}=W_{1}=f\equiv 0$, $\sigma\equiv 1$,
$N=1-2M/r+1/r^{2}\equiv(r-r_{+})(r-r_{-})/r^{2}$, $M$ being  the
ADM mass. We specify ${\cal D}(H)$ as before and choose the set
$\{ \Psi_{k}\} \subset {\cal D}(H)$ parameterized by $\alpha_{k}=0$,
$\xi_{k}=(r-r_{+})/r^{2}h_{k}$.
Repeating the above procedure, one arrives at
\be
\langle\Psi_{k}|H|\Psi_{k}\rangle=
\int_{-\infty}^{\infty}dr_{\ast} \left((\xi_{k})'^{2}_{\ast}-
\frac{N}{r^{2}}\xi_{k}^{2} \right)=
-\frac{7r_{+}+5r_{-}}{420r_{+}^{4}}+I(k),            \label{26}
\ee
where $I(k)\rightarrow 0$ with growing $k$.
This shows  that Abelian EYM  black  holes  are
unstable  with  respect   to non--Abelian fluctuations
(see also \cite{loh}).

\section{Discussion}

Our analysis reveals  the   existence   of  at   least   one
{\em odd--parity}~ negative mode for all known non-Abelian
EYM black hole solutions,
indicating in particular that all of them are unstable.
For the uncharged $SU(2)$ solutions, taking
into account the results of the analysis  by  Straumann   and   Zhou
\cite{SZ}, we therefore conclude  that  each  $(n,r_{H})$ black
hole has at least $n+1$ unstable modes -- $n$ in the even-parity
sector and at least one odd-parity negative mode.
Our   analysis   is   equally
valid  in the regular case, where  such an odd-parity
mode has precisely the  same  meaning  as negative mode of the
electroweak sphaleron solution \cite{guys}, \cite{kunz}.
In this sense, this mode is  fairly interesting and,
from the physical point of view, quite typical. In the regular case,
it had been precisely the existence of this mode which allowed
us  to  suggest a sphaleron  interpretation  for  the  BK   solutions
\cite{spha}. On the other hand, the existence of the {\em even-parity}~
sphaleron negative modes is a rather  peculiar  phenomenon  which  is
present in the EYM theory \cite{sphal}.

For  black   holes,   as  has  already      been    mentioned
\cite{old}, the sphaleron interpretation  is  not  so transparent. By
definition, a sphaleron  is   a  static  solution "sitting"  at   the
top     of    a potential      barrier     separating   topologically
distinct YM vacua, whereas  in  the  black  hole  case there are  no
pure vacuum states because of  the  finite   temperature  associated
with  the  event   horizon   (except   for    the    extreme    case
\cite{charge}, \cite{hajicek}). The existence of the odd--parity
negative
modes in the black hole case shows, however, that the structure of the
energy surface in the vicinity of the EYM black hole solutions in
function space is similar to that for the regular BK objects. This
suggests one to think of them as ``black holes inside EYM sphalerons''.
An interesting open issue is the  {\em  exact}
number of the odd--parity modes both for regular and  black  hole EYM
solutions. Investigation of this requires
a further numerical work.

\section*{Acknowledgments}
We  would  like  to  thank Ruth Durrer for careful reading of the
manuscript and Professor  N.  Straumann  for reading of the
manuscript and valuable discussions. The work of MSV
was supported by the Swiss National Science Foundation. The work  of
DVG was supported by the Russian Foundation for Fundamental Research
grant 93--02--16977, the ISF grant M79000, and by CONACyT (Mexico).

\end{document}